\documentclass[centertags,12pt]{amsart}
\usepackage{latexsym}
\usepackage{amssymb}
\usepackage{graphics}

\numberwithin{equation}{section}

\makeatletter
\renewcommand{\subsection}{\@startsection
{subsection}{2}{0mm}{\baselineskip}{-0.25cm}
{\normalfont\normalsize\em}}
\makeatother

\newtheorem{theorem}{Theorem}[section]
\newtheorem{proposition}[theorem]{Proposition}
\newtheorem{corollary}[theorem]{Corollary}
\newtheorem{lemma}[theorem]{Lemma}

{\theoremstyle{definition}
\newtheorem{definition}[theorem]{Definition}
\newtheorem{example}[theorem]{Example}

}

\theoremstyle{remark}
\newtheorem{remark}[theorem]{Remark}

\def\F{\mathbb F}
\def\N{\mathbb N}

\def\1{\mathbf 1}

\def\R{\mathbf R}

\def\cB{\mathcal B}

\def\cL{\mathcal L}
\def\cM{\mathcal M}
\def\cO{\mathcal O}

\def\cX{\mathcal X}

\def\cU{\mathcal U}

\def\Gaps{{\rm Gaps}}

\def\div{{\rm div}}

\def\dim{{\rm dim}}
\def\deg{{\rm deg}}

\def\by{\mathbf y}
\def\bh{\mathbf h}

\begin{document}

\author{C. Carvalho}
\author{C. Munuera}
\author{E. Silva}
\author{F. Torres}

\title[Near Orders and Codes]{Near Orders and Codes}

\address{Faculdade de Matem\'atica, Universidade Federal de
Uberl\^andia, Av. J.N. de \'Avila 2160, Uberl\^andia, 38408-100,
Uberl\^andia, MG-Brazil.}

\email{cicero@ufu.br}

\address{Department of Applied Mathematics, University of Valladolid
(ETS Arquitectura) 47014 Valladolid, Castilla, Spain.}

\email{cmunuera@modulor.arq.uva.es}

\address{Faculdade de Matem\'atica, Universidade Federal de
Uberl\^andia, Av. J.N. de \'Avila 2160, Uberl\^andia, 38408-100,
Uberl\^andia, MG-Brazil}

\email{ercilio@ufu.br}

\address{IMECC-UNICAMP, Cx.P. 6065, 13083-970, Campinas SP-Brazil.}

\email{ftorres@ime.unicamp.br}


\thanks{{\em 2000 Math. Subj. Class.}: 94B27; 14G50}

\thanks{The work of C. Carvalho was partially supported by FAPEMIG,
grant CEX 605/05; research done under a joint project of the
Millennium Institute for the Global Advancement of Brazilian
Mathematics (IM-AGIMB) and Universidade Federal de Uberl\^andia (UFU)}

\thanks{The work of C. Munuera was supported by the ``Junta de
Castilla y Le\'on", Espa\~na, under Grant VA020-02}

\thanks{The work of E. Silva was partially supported by FAPEMIG; this 
paper is based on his Ph.D dissertation \cite{ercilio} done at 
IMECC-UNICAMP, SP-Brazil}

\thanks{The work of F. Torres was supported by CNPq-Brazil (306676/03-6) 
and PRONEX (66.2408/96-9)}


   \begin{abstract} H\o holdt, van Lint and Pellikaan 
used order functions to construct codes by means of Linear Algebra and
Semigroup Theory only. However, Geometric Goppa codes that can be
represented by this method are mainly those based on just one point.
In this paper we introduce the concept of near order function with the
aim of generalize this approach 
in such a way that a of 
wider family of Geometric Goppa codes can be studied on a more 
elementary setting.
   \end{abstract}

\maketitle

   \section{Introduction}\label{s1}

Geometric Goppa codes (or GG codes, for short) were constructed by
Goppa \cite{goppa1}, \cite{goppa2} based on a curve $\cX$ over a
finite field $\F$, and two $\F$-rational divisors $D$ and $G$ on
$\cX$. Here, by a {\em curve} we mean a projective, geometrically
irreducible, non-singular algebraic curve. Usually the divisors $D$
and $G$ are chosen as
   \begin{itemize}
\item $D=P_1+\ldots+P_n$;
\item $G=\alpha_1Q_1+\ldots+\alpha_\ell Q_\ell$,
   \end{itemize}
where the $P_i$'s and $Q_j$'s are pairwise different $\F$-rational
points of $\cX$. Then, there are two GG codes associated to the triple
$(\cX,D,G)$, defined as the images $C_{\cL}=C_\cL(\cX,D,G)$ and
$C_{\Omega}=C_\Omega(\cX,D,G)$ of the maps
    $$
ev:f\in \cL(G)\mapsto (f(P_1),\ldots,f(P_n))\in \F^n\quad \text{and}
    $$
    $$
res:\omega\in \Omega(G-D)\mapsto
(\text{res}_{P_1}(\omega),\ldots,\text{res}_{P_n}(\omega))\in \F^n
    $$
respectively. According to the residue theorem, these codes are dual to 
the other, $C_{\cL}=C_{\Omega}^{\perp}$, hence both constructions provide 
the same family of codes. Bounds on the dimension and minimum distance of 
such codes are available from their definition, as they satisfy 
$k=\ell(G)-\ell(G-D)$, $d\geq n-\deg(G)$ for $C_{\cL}$ and 
$k=i(G-D)-i(G)$, $d\geq \deg(G)-2\gamma+2$ for $C_{\Omega}$ (where 
$\gamma$ is the genus of $\cX$). Soon after its introduction, GG codes 
became a very important tool in Coding Theory; for example, Tsfasman, 
Vladut and Zink \cite{tsf-vla-z} showed that the Varshamov-Gilbert bound 
can be attained by using these codes. The way of dealing with the 
dimension and minimum distance of $C$ is via the Riemann-Roch theorem; in 
particular one needs to compute the genus of the underlying curve which 
may be a difficult task. Thus it will be of interest to construct and 
manage GG codes by using ``elementary methods" only. An important step in 
this direction was given by H\o holdt, van Lint and Pellikaan 
\cite{h-vl-p} (see also \cite{feng}), who used order functions (see 
Section \ref{s2.2}) to construct codes from an $\F$-algebra $\R$. Order 
functions and the obtained codes have benn studied in detail by 
Pellikaan, Geil and other authors (see \cite{ord1}, \cite{ord2}). This 
technique allows us to do mainly with ``one-point GG" codes --that is to 
say, when $\ell=1$ in the definition of divisor $G$ above--. The 
objective of this paper is to introduce and study a wider class of 
``order-like" functions (the called 
{\em near order} functions; see Section \ref{s3}) in such a way that more 
GG codes could be represented by those elementary methods.

   \section{Background}\label{s2}

\subsection{Weierstrass Semigroups and Geometric Goppa Codes}\label{s2.1}

Let $\cX$ be a curve over a finite field $\F$. For a point $P\in\cX$,
let $\cO_P$ and $v_P$ denote the local ring and valuation of $\cX$ at
$P$ respectively. Following \cite{h-vl-p}, we consider the
$\F$-algebra
   $$
\R=\R(Q_1,\ldots,Q_\ell):=\bigcap_{R\neq Q_1,\ldots,Q_\ell}\cO_R\, ,
   $$
where the $Q_i$'s are as in Section \ref{s1}; we shall consider also the
Weierstrass semigroup of $\cX$ at $Q_1,\ldots,Q_\ell$, namely
   \begin{align*}
H & = H(Q_1,\ldots,Q_\ell)\\
{} & =\{(\beta_1,\ldots,\beta_\ell)\in\N_0^\ell: \text{
there exists  $f\in\R$ with $\div_\infty(f)=\beta_1Q_1+\ldots+\beta_\ell
Q_\ell$}\}\, .
   \end{align*}
These semigroups have been intensively studied in connection with Coding 
Theory; see for example \cite{ct}, \cite{garcia-lax1}, \cite{garcia-lax2}, 
\cite{homma}, \cite{homma-kim1}, \cite{homma-kim2} \cite{homma-kim3} 
\cite{homma-kim4}, \cite{kim}, \cite{gretchen3}, \cite{gretchen1}, 
\cite{gretchen2}, \cite{gretchen-survey}. The relationship between $\R$ 
and $H$ above suggests that Goppa codes can be represented by elementary 
means. As was already mentioned, this was noticed in \cite{h-vl-p} for the 
case $\ell=1$ (see also \cite{matsumoto}).

    \subsection{Order Functions}\label{s2.2}

Our reference in this section is the paper \cite{h-vl-p}. Let $\R$ be
an $\F$-algebra. A function $\rho:\R\to\N_0\cup\{-\infty\}$ is called
an {\em order} function if the following properties
    \begin{enumerate}
\item[\rm(O0)] $\rho(f)=-\infty$ if and only if $f=0$;
\item[\rm(O1)] $\rho(\lambda f)=\rho(f)$ for all $\lambda\in \F^*$;
\item[\rm(O2)] $\rho(f+g)\leq \max\{\rho(f), \rho(g)\}$;
\item[\rm(O3)] If $\rho(f)<\rho(g)$ and $h\neq 0$, then
$\rho(fh)<\rho(gh)$; and
\item[\rm(O4)] If $\rho(f)=\rho(g)\neq -\infty$, then there
exists  $\lambda \in \F^*$ such that $\rho(f-\lambda g)<\rho(g)$,
     \end{enumerate}
are satisfied for all $f,g,h\in\R$. If in addition
    \begin{enumerate}
\item[\rm(O5)] $\rho(fg)=\rho(f)+\rho(g)$,
    \end{enumerate}
then $\rho$ is called a {\em weight} function. We collect some
properties of order functions.
    \begin{lemma}{\rm (\cite[Lemma 3.9]{h-vl-p})}\label{lemma2.21}
With notation as above$:$
   \begin{enumerate}
\item If $\rho(f)=\rho(g),$ then $\rho(fh)=\rho(gh)$ for all $h\in\R;$
\item If $f\in\R\setminus\{0\}$, then $\rho(1)\leq \rho(f);$
\item $\F^*=\{f\in\R: \rho(f)=\rho(1)\};$
\item If $f\neq 0$, $g\neq 0$ and $\rho(f)=\rho(g),$ then there exists
a unique nonzero
$\lambda\in \F$ such that $\rho(f-\lambda g)<\rho(f);$
\item If $\rho(f)\neq\rho(g),$ then $\rho(f+g)= \max\{\rho(f),
\rho(g)\}.$
   \end{enumerate}
   \end{lemma}
   \begin{remark}\label{rem2.21} According to the lemma above, the
$\F$-algebra $\R$ splits as $\R=\cM\cup\cU$, where
   $$
\cM=\{f\in\R:\rho(f)>\rho(1)\}\, , \quad\text{and}\quad
\cU=\{f\in\R:\rho(f)\leq \rho(1)\}\, .
   $$
As a matter of fact, $\cU^*:=\cU\setminus\{0\}=\{f\in
\R\setminus\{0\}:\rho(f)=\rho(1)\}=\F^*$.
   \end{remark}

    \section{Near Order Functions}\label{s3}

In this section we study a ``weak" version
of the concept of order and weight function discussed in Section
\ref{s2}. The starting point for our discussion is Remark
\ref{rem2.21}.

   \subsection{Near Order Functions}\label{s3.1}

Let $\R$ be an $\F$-algebra and let $\rho:\R\to\N_0\cup\{-\infty\}$ be
a function with $\rho(0)=-\infty$. Associated to $\rho$ we can consider
the sets:
   \begin{align*}
\cU=\cU_\rho & := \{f\in \R :
\rho(f)\leq \rho(1)\}\, ,\\ 
\cU^*=\cU_\rho^* &:=  \cU\setminus\{0\}\, ,\\
\cM=\cM_\rho &:=  \{f\in \R: \rho(f)>\rho(1)\}\, .
    \end{align*}
In addition, let $\cU=\cU_\rho:=\cU^*\cup\{0\}$. 
We say that $\rho$ is
a {\em near order} function (or {\em n-order} function, for short) if
properties
   \begin{enumerate}
\item[\rm(N0)] $\rho(f)=-\infty$ if and only if $f=0$;
\item[\rm(N1)] $\rho(\lambda f)=\rho(f)$ for all $\lambda\in \F^*$;
\item[\rm(N2)] $\rho(f+g)\leq \max\{\rho(f), \rho(g)\}$;
   \end{enumerate}
similar to the corresponding concerning order functions hold true, and for
$f,g,h\in\R$ we have:
   \begin{enumerate}
\item[\rm(N3)] If $\rho(f)<\rho(g)$ then $\rho(fh)\leq \rho(gh)$.
Furthermore, if $h\in \cM$, then $\rho(fh)<\rho(gh)$;
\item[\rm(N4)] If $\rho(f)=\rho(g)$ with $f,g\in \cM$, then there
exists $\lambda\in \F^*$ such that $\rho(f-\lambda g)<\rho(f)$.
    \end{enumerate}
Clearly an order function is also a n-order (cf. Remark \ref{rem2.21}).
We can also construct n-orders functions on $\R$ which are not orders
functions.
    \begin{example}\label{example3.11}
(a) Let $\rho(0)=-\infty$ and for $f\in\R\setminus\{0\}$ put
$\rho(f):=c\in\N_0$ (constant). Here $\cM=\emptyset$ and $\cU=\R$, so
$\rho$ is trivially a n-order on $\R$ which is not an order (it is an
order function if and only if $\R=\F$).

(b) Fix $g\in \R\setminus \F$ and define $\rho(f)=-\infty$ if and only if
$f=0$; $\rho(f):=0$ if $f\in \langle g\rangle, f\neq 0$; $\rho(f)=1$
otherwise. Then $\rho$ is a n-order function with $\rho(1)=1$ and
$\cU=\R$.
    \end{example}
The examples above shows the existence of n-order functions on an
arbitrary $\F$-algebra. Note that in both cases it holds that $\cU=\R$
hence $\cM=\emptyset$. N-orders verifying this condition will be
called {\em trivial}. For non-trivial n-orders both sets $\cM$ and $\rho(\cM)$
have infinitely many elements. This is a consequence of (N3), since $\rho(1)<\rho(f)$ implies
$\rho(f^i)<\rho(f^{i+1})$. 
An example of a nontrivial n-order is the following.

    \begin{example}\label{example3.12} Let $\R=\F[X,Y]/(XY-1)=\F[x,y]$
with $x, y$ being the class of $X$ and $Y$ respectively. Every
$f\in\R$ admits a unique decomposition of type
$f=f_1(x)+f_2(y)$, where $f_1, f_2\in \F[T]$ with
$f_2(0)=0$. It is known that $\R$ does not admit any order
function, \cite[Ex. 3.11]{h-vl-p}. However, $\R$ admits a
non-trivial  n-order function, namely
    $$
\rho(f):=
  \begin{cases}
-\infty   & \text{if $f=0$},\\
      0   & \text{if $f_1\neq 0$ and $f_2=0$},\\
\deg(f_2) & \text{if $f_2\neq 0$.}
   \end{cases}
   $$
Here $\rho(1)=0$, $\cM=\{f_1(x)+f_2(y): \text{$f_2\in \F[t]$, $f_2\neq
0$ and $f_2(0)=0$}\}$, $\cU=\{f_1(x): f_1\in \F[t]\}$; an
straightforward computation shows that $\rho$ is in fact a n-order
function.
   \end{example}
The relation between orders and n-orders is clarified by the
following result, which complements Remark \ref{rem2.21}.
    \begin{lemma}\label{lemma3.11}
Let $\rho:\R\to\N_0\cup\{-\infty\}$ be a function defined on a
$\F$-algebra $\R$. Let $\cU=\cU_\rho$ be the set of elements $f\in\R$
with $\rho(f)\leq \rho(1).$ Then the following statements are
equivalent:
   \begin{enumerate}
\item $\rho$ is an order$;$
\item $\rho$ is a n-order and $\cU=\F.$
   \end{enumerate}
   \end{lemma}
Note that, as a consequence of property (N1), for any n-order on $\R$
it holds that $\F\subseteq\cU$. The above lemma shows that equality
holds just for orders. On the other hand, it was noticed in
\cite[Prop. 3.10]{h-vl-p} that any $\F$-algebra equipped with an order
function is an integral domain but the inverse statement is false; cf.
Example \ref{example3.12}. We stress the fact that any $\F$ algebra
can be equipped with a n-order function; cf. Examples
\ref{example3.11}, \ref{example3.12}.

     \begin{lemma}\label{lemma3.12} Let $\R$ be an $\F$-algebra
and $\rho$ a n-order on $\R$. Then the set $\cM_\rho$ does not contain zero
divisors.
     \end{lemma}
     \begin{proof} Let $g\in \R\setminus\{0\}$ and $f\in\cM$. Since
$\rho(1)<\rho(f)$ it holds that $\rho(g)\leq \rho(fg)$ by Axiom
$(N3)$. Hence $fg\neq 0$.
     \end{proof}
Let us see one more example. As said in the Introduction, our purpose
is to manage Goppa codes over more that one point by means of
``order-like" functions. This example shows a way to obtain n-order
functions from points on curves (cf. Section \ref{s6}).
   \begin{example}\label{example3.13} Let $\cX$ be a curve over a
finite field $\F$. Let $Q_1,\ldots, Q_\ell$ be pairwise different
$\F$-rational points of $\cX$ and $\R=\R(Q_1,\ldots, Q_\ell)$ the
algebra defined in Section \ref{s2.1}.  For each point $Q_i$, define
the function $\rho_i=\rho_{Q_i}:\R\to\N_0\cup\{-\infty\}$ by
$\rho_i(f)=-\infty$ if and only if $f=0$, and
   $$
\rho_i(f)=\begin{cases}
   0  &  \text{if $v_{Q_i}(f)\geq 0$},\\
   -v_{Q_i}(f) & \text{if $v_{Q_i}(f)<0$}.
   \end{cases}
   $$
Then $\rho_i(1)=0$ and hence $\cU_i^*=\{f\in\R^*: v_{Q_i}(f)\geq 0\}$. As a
consequence of properties regarding valuation maps,
$\rho_i$ is indeed a n-order function (and in fact a n-weight as
we shall define it later).

Note that in the one-point case ($\ell=1$) it holds that
   $$
H(Q_1)=\{-v_{Q_1}(f): f\in \R^* \}\, .
   $$
In the multiple-point case ($\ell>1$) we must use the functions $\rho_i$'s 
instead of the valuations $-v_{Q_i}$'s in order to describe the 
Weierstrass semigroup; indeed,
  $$
H(Q_1,\ldots,Q_\ell)=\{(\rho_1(f),\ldots,\rho_\ell(f)): f\in \R^*\}\, .
   $$
This fact gives a motivation to define the concept of near order.
   \end{example}
Now we subsume further properties of n-order functions that
are similar to those of order functions (cf. \cite[Lemma 3.9]{h-vl-p}).
    \begin{lemma}\label{lemma3.13} Let $\rho$ be a n-order function
on a $\F$-algebra $\R$. The following statements hold:
   \begin{enumerate}
\item[\rm(1)] If $f,g,h\in\cM_\rho$ and $\rho(f)=\rho(g),$ then
$\rho(fh)=\rho(gh);$
\item[\rm(2)] The element $\lambda$ in Axiom $(N4)$ is unique;
\item[\rm(3)] If $\rho(f)\neq\rho(g),$ then $\rho(f+g)= \max\{\rho(f),
\rho(g)\}$.
   \end{enumerate}
   \end{lemma}
    \begin{proof} Similar to the proof of
\cite[Lemma 3.9(1),(2),(4)]{h-vl-p}.
    \end{proof}

   \subsection{Normalized Near Orders and Near Weights
Functions}\label{s3.2}

Let $\rho$ be a n-order function on $\R$. As we shall see in the
forthcoming sections, we will be interested in the value of $\rho(f)$
when $f\in\cM$ but not when $f\in\cU$. Thus we can consider the {\em
normalization} of $\rho$
as the function $\tilde{\rho}$ defined as $\tilde{\rho}(0)=-\infty$
and for $f\neq 0$
   $$
\tilde{\rho}(f)=\left\{\begin{array}{ll}
 0         & \mbox{if $f\in\cU_{\rho}$; } \\
 \rho(f)/d & \mbox{if $f\in\cM_{\rho}$,}
\end{array} \right.
  $$
where $d=\gcd (\rho(\cM))$. It is clear that $\tilde{\rho}$ is also
a n-order function, $\cM_{\tilde\rho}=\cM_\rho$
and $\cU_{\tilde\rho}=\cU_\rho$. The n-order function $\rho$ is
said to be {\em normal} if $\rho=\tilde{\rho}$. In what follows, all
the n-orders functions we consider will be understood as normal.

A (normal) n-order function $\rho$ is called a {\em near weight} (or
{\em n-weight}, for short) if it verifies the supplementary condition
   \begin{enumerate}
\item[(N5)] $\rho(fg)\leq \rho(f)+\rho(g)$. If $f,g\in\cM$, then
   equality holds.
   \end{enumerate}
Two interesting properties of n-weights arise at once from its definition
     \begin{proposition}\label{prop3.20}
Let $\rho$ be a n-weight function on the $\F$-algebra $\R$. Then
   \begin{enumerate}
\item[\rm(1)] the set $\rho(\R\setminus\{0\})$ is a
numerical semigroup of finite genus;
\item[\rm(2)] the set $\cU_\rho$ is closed under
product and so it is a subalgebra of $\R$.
    \end{enumerate}
   \end{proposition}
Next, motivated by Proposition 3.12 and Theorem 3.14 in \cite{h-vl-p},
we point out a relation between n-order functions $\rho$ on $\R$ and
subspaces of $\R$. Set $\rho(\R\setminus\{0\})=\{ 0=\rho_0<\rho_1<\rho_2<\ldots \}$ and
   \begin{itemize}
\item For $i\in\N_0$, $L_i:=\{f\in\R:\rho(f)\leq \rho_i\}$;
\item For $f\neq 0$ define $\iota(f)$ as being the least
non-negative integer $\ell$ such
that $f\in L_\ell$;
\item For $i,j\in\N_0,$ $\ell(i,j):=\max\{\iota
(fg): \text{$f\in L_i$ and  $g\in L_j$}\}$.
   \end{itemize}
   \begin{proposition}\label{prop3.21} Let $\rho$
be a n-order function on a $\F$-algebra $\R$ whose set of non-units
is not empty$.$ Then the following statements hold true$:$
    \begin{enumerate}
\item $(L_i)$ is an increasing sequence of vector
subspaces of $\R$ such that$:$
  \begin{enumerate}
\item $\F\subseteq L_0;$
\item $\dim(L_{i+1})=\dim (L_i)+1;$
\item $\cup_i L_i=\R;$
   \end{enumerate}
\item $\ell(i,j)=\ell(j,i),$ and for all $i\in\N_0:$
    \begin{enumerate}
 \item If $j\geq 1,$ then $\ell(i,j)<\ell(i+1,j);$
 \item If $j=0,$ then $\ell(i,0)\leq\ell(i+1,0);$
    \end{enumerate}
\item If $\rho$ is a n-weight function$,$ then
$\rho_{\ell(i,j)}\leq\rho_i+\rho_j$ for $i,j\in\N_0.$ If $i,j\geq 1$
the equality holds$.$
    \end{enumerate}
    \end{proposition}
     \begin{proof} (1) The fact that the $L_i$'s are vector spaces
follows from properties (N0),(N1) and (N2) of n-order functions.
(1.a) was already noted. (1.b) holds as a consequence of properties
(N2) and  (N4). Statement (1.c) is obvious; (2) is
a direct consequence of property (N3); (3) follows from (N5).
     \end{proof}
     \begin{remark}\label{rem3.22} As doing in \cite{h-vl-p}, the
above proposition can be partially written in terms of functions
instead of subspaces. Indeed, for $i\in\N$, let $f_i\in
L_{i}\setminus L_{i-1}$. Then
     \begin{enumerate}
\item $\iota(f_i)=i$ and $\rho(f_i)=\rho_i$;
\item The set $(f_i)$ is linearly independent and $\R=\cU\oplus\langle
f_1,f_2\ldots\rangle$.
\item For $i,j\in\N$, $\ell(i,j)=\iota(f_if_j)$.
     \end{enumerate}
     \end{remark}
Conversely, we will prove that certain sequences of subspaces of $\R$
defines a n-order function on $\R$. Let $L_0\subseteq
L_1\subseteq\ldots$ be an increasing sequence of vector subspaces of
$\R$ verifying the conditions:
   \begin{enumerate}
\item[\rm(1a)] $\F\subseteq L_0$;
\item[\rm(1b)] $\dim (L_{i+1})=\dim (L_i)+1$;
\item[\rm(1c)] $\cup_i L_i=\R$.
   \end{enumerate}
Let $0=\rho_0<\rho_1<\ldots$ be a sequence of positive integers whose
cardinality is the same as the sequence $(L_i)$ and such that
$\gcd(\rho_i)=1$. For $f\neq 0$ define $\iota(f)$ as being the least
non-negative integer $\ell$ such that $f\in L_\ell$. For $i,j\in\N_0$,
set
   $$
\ell(i,j):=\max\{\iota (fg): \text{$f\in L_i$ and $g\in L_j$}\}\, .
   $$
The following proposition arises:
    \begin{proposition}\label{prop3.22} Notation as above. Let
$\rho:\R\to \N_0\cup\{-\infty\}$ be the
function defined by $\rho(0)=-\infty$ and $\rho(f)=\rho_{\iota(f)}$
for $f\neq 0.$ If the following two conditions$:$
    \begin{enumerate}
\item[\rm(2a)] If $j\geq 1,$ then $\ell(i,j)<\ell(i+1,j),$
\item[\rm(2b)] If $j=0,$ then $\ell(i,0)\leq\ell(i+1,0)$
    \end{enumerate}
hold$,$ then $\rho$ is a n-order function$;$ if$,$ in addition$,$
$\rho_{\ell(i,j)}\leq\rho_i+\rho_j$ with equality if $i,j\geq 1,$ then
$\rho$ is a n-weight function$.$ In both cases, $L_0\setminus\{0\}$ is
the set of unities of $\R.$
    \end{proposition}

   \section{Well-Agreeing n-weights}\label{s5}

The subject matter of this section can be applied to finitely many
n-weight functions. However, for simplicity we shall
consider the case of just two n-weights. Then, let $\rho,\sigma$ be
two n-weights on $\R$. We consider the following subsemigroup of
$(\N^2_0,+)$:
   $$
H(\rho,\sigma)=\{(\rho(f),\sigma(f)):f\in\R\setminus\{0\}\}\, .
   $$
It can have, or have not, a finite genus.
    \begin{proposition}\label{prop5.1}
If $H(\rho,\sigma)$ has a finite genus$,$ then
$\rho(\R\setminus\{0\})=\N_0$ and $\sigma(\R\setminus\{0\})=\N_0.$
     \end{proposition}
     \begin{proof}
Let $n\in\N_0$. The set $\{(n,m)\notin H(\rho,\sigma): m\in\N_0
\}$ is finite, hence $n\in \rho(\R)$. Analogously for $\sigma$.
    \end{proof}
In what follows we shall assume that $H(\rho,\sigma)$ has a finite
genus. As said before, both $\cU_{\rho}$ and $\cU_{\sigma}$ are 
subalgebras of $\R$. Then, the sets
  \begin{align*}
H(\sigma) &:=\sigma(\cU_{\rho}^*)=\{m:(0,m)\in H(\rho,\sigma)\}\quad  
\text{and} \\ 
H(\rho)&:=\rho(\cU_{\sigma}^*)=\{m:(m,0)\in H(\rho,\sigma)\} 
   \end{align*}
are numerical semigroups. Write
   $
H(\sigma)=\{ 0=m_0<m_1<m_2<\ldots \} .
  $
  \begin{lemma}\label{lemma5.1} 
The semigroups $H(\sigma)$ and $H(\rho)$ have
at most the genus of $H(\rho,\sigma)$.
   \end{lemma}
   \begin{proof} Note that 
$H(\sigma)=\sigma(\cU_{\rho}\cap\cM_{\sigma})\cup\{0\}$. Now 
if $\ell$ is a gap of $H(\sigma)$, then $(0,\ell)$ is a gap 
of $H(\rho,\sigma)$. The same argument for $H(\rho)$.
   \end{proof}
    \begin{proposition}\label{prop5.2} Let $f_0=1$ and for $i\in\N$
take functions $f_i\in \R,$ $g_i\in\cU_{{\rho}}$ such that
$\rho(f_i)=i,$ $\sigma(g_i)=m_i.$ Set
$\cB:=\{f_i:i\in\N_0\}\cup\{g_j:j\in\N\}\subseteq \R.$ If
$\cU_\rho\cap\cU_\sigma=\F,$ then $\cB$ is a basis of $\R$ as a
$\F$-vector space$.$
   \end{proposition}
    \begin{proof}  We first show that $\cB$ is a linearly independent set.
If $\lambda_0f_0+\ldots+\lambda_rf_r=\mu_1g_1+\ldots+\mu_sg_s$, then
   $$
\rho(\lambda_0f_0+\ldots+\lambda_rf_r)=\rho(\mu_1g_1+\ldots+\mu_sg_s)
= 0\, ,
   $$
by (N2) and $g_j\in\cU_\rho$. Then $\lambda_i=0$ for $i\geq 1$ by
\ref{lemma3.13}(3),
and so
   $$
-\lambda_0+\mu_1g_1+\ldots +\mu_sg_s=0\, .
   $$
As above it follows that $\mu_1=\ldots=\mu_s=0$ and so $\cB$ is in
fact a linearly independent set.

We show next that $\cB$ generates $\R$. Let $h\in \R$ such that
$\rho(h)=i\in\N_0$. By applying iteratively (N4), there
exist elements $\lambda_1,\ldots,\lambda_r\in \F$ such that
   $$
\tilde h:=h-\lambda_1f_1-\ldots-\lambda_rf_r\in\cU_\rho\, .
   $$
Let $\sigma(\tilde h)=m_s=\sigma(g_s)$. Arguing as above, we
find elements $\beta_1,\ldots,\beta_s\in\F$ so that
   $$
\tilde h -\beta_1g_1-\ldots-\beta_sg_s\in\cU_\sigma\, .
   $$
The proof now follows by the hypothesis $\cU_\rho\cap\cU_\sigma=\F$.
    \end{proof}
   \begin{definition}\label{definition5.1}
We say that the n-weights $\rho$ and $\sigma$ {\em agree
well} if the semigroup $H(\rho,\sigma)$ has a finite genus and
$\cU_{\rho}\cap\cU_{\sigma}=\F$.
    \end{definition}
   \begin{example}\label{example5.1} (Continuation of Example
\ref{example3.13}) Let $\cX$ be a curve of genus $\gamma$ over $\F$
and let $Q_1,Q_2$ be two rational points. Let $\R=\R(Q_1,Q_2)$ and
$\rho,\sigma$ be the n-weights associated to the points $Q_1,Q_2$
respectively. Then $H(\rho,\sigma)$ is just the Weierstrass semigroup
at $Q_1,Q_2$, $H(\rho,\sigma)=H(Q_1,Q_2)$. By the Riemann-Roch
theorem this semigroup has finite genus. Furthermore
since $\cU_{\rho}$ (resp. $\cU_{\sigma}$) is the set of rational functions
having poles only at $Q_2$ (resp. at $Q_1$), then
$\cU_{\rho}\cap\cU_{\sigma}=\F$, hence $\rho$ and $\sigma$ agree well.
Moreover, it is easy to see that $H(\rho)=H(Q_1)$ and
$H(\sigma)=H(Q_2)$. In this case (n-weights associated to
points on a curve), both semigroups have the same genus, $\gamma$. As
we shall see next, this is also true for general well agreeing
n-weights; see Corollary \ref{cor5.1}.
    \end{example}
If the n-weights $\rho$ and $\sigma$ agree well, then the
functions $f_i$ in the basis $\cB$  can be taken in such a way that (cf. 
\cite{kim})
    \begin{equation}\label{eq5.1}
\sigma(f_i)=\min\{\sigma(f): \text{$f\in\R$ and $\rho(f)=i$}\}\, .
     \end{equation}
    \begin{definition}\label{definition5.2}
A basis with the property above will be called {\em good} (with respect to 
the n-weights $\rho$ and $\sigma$).
   \end{definition}
The next proposition and its corollary states some properties of good 
basis.
   \begin{proposition}\label{propgood} Let $\rho$ and $\sigma$ be two
well agreeing n-weights on $\R$ and let
$\cB=\{f_i:i\in\N_0\}\cup\{g_j:j\in\N\}$ be a good basis. Then
  \begin{enumerate}
\item For all $i=0,1,\dots,$ either $\sigma(f_i)=0$
or $\sigma(f_i)$ is a gap of $H(\sigma);$
\item Conversely$,$ for every gap $m$ of $H(\sigma)$ there
exists exactly one index $i$ such that $\sigma(f_i)=m;$
\item $\sigma(f_i)=0$ if and only if $i$ is a nongap of
$H(\rho)$.
    \end{enumerate}
    \end{proposition}
   \begin{proof} (1) Suppose that
$\sigma(f_i)=m_j\in H(\sigma)$,  $m_j\neq 0$; then
$\sigma(f_i)=\sigma(g_j)$ and by
(N4) there exists $\mu_j\in\F^*$ such that $\sigma(f_i-\mu_jg_j)<m_j$.
Proceeding iteratively in this way we find a $\tilde f_i:=f_i+
\sum_j\mu_jg_j$ such that $\sigma(\tilde f_i)\in {\rm
Gaps}(H(\sigma))\cup\{0\}$. Since $\rho(\tilde f_i)=\rho(f_i)=i$ by
Lemma \ref{lemma2.21}(5), the proof is complete. 

(2) Let $t\in\N$ be a gap of $H(\sigma)$. Let us prove first
that there are at most one $r$ such that $\sigma(f_r)=t$. If, on the
contrary, $\sigma(f_i)=\sigma(f_j)=t$ for some $i>j$, then there is
a $\lambda\in\F^*$ such that $\sigma(f_i-\lambda f_j)<m$. Then
$\tilde f_i=f_i-\lambda f_j$ verifies $\rho(\tilde f_i)=i$,
contradicting the defining property of the function $f_i$. Let us
prove now that there is an index $r$ such that $\sigma(f_r)=t$. From
\ref{prop5.1} there is $h\in\R$ such that $\sigma(h)=m$. Write
   $$
h=\sum_{i\in I} \lambda_if_i+\sum_{j\in J} \mu_jg_j\, .
   $$
with $I\subseteq\N_0, J\subset\N$ and $\lambda_i,\mu_j\neq 0$. Since all
the elements in the family $\{ \sigma(f_i): i\in I, \sigma(f_i)\neq 0
\}$ are pairwise distinct gaps of $H(\sigma)$, and all the
elements in the family $\{ \sigma(g_j): j\in J \}$ are pairwise
distinct nongaps of $H(\sigma)$, according to the properties
of n-weights, we conclude that
  $$
t=\sigma(h)=\max ( \{ \sigma(f_i): i\in I\} \cup\{\sigma(g_j):
j\in J\} )=\sigma(f_r)
  $$
for some $r$ (because $m$ is a gap). 

(3) $\sigma(f_i)=0$ if and only if there exists $f\in\cU_{\sigma}$
with $\rho(f)=i$, that is, if and only if $i$ is a nongap of $H(\rho)$.
\end{proof}
    \begin{corollary}\label{cor5.1} Let $\rho$ and $\sigma$ be two
well agreeing n-weights on $\R$ and let $\cB$ be a good basis. Then
   \begin{enumerate}
\item $\sigma(f_i)$ is a gap of $H(\sigma)$ 
if and only if $i$ is a gap of $H(\rho)$. In particular,
both semigroups have equal genus.
  \item $\sigma(f_i)=0$ except for finitely many $i$'s; for all
$i,$ $\sigma(f_i)\leq\Lambda_{\sigma},$
where $\Lambda_{\sigma}$ is the largest gap of $H(\sigma)$.
   \end{enumerate}
    \end{corollary}
Well agreeing n-weights and good basis can be used to construct codes from 
$\R$, as we shall see in the next Section.

   \section{The Codes and a Bound on the Minimum Distance}\label{s6} 

   \subsection{N-order Codes}\label{s6.1}

Let $\rho,\sigma$ be two well agreeing n-weights on a $\F$-algebra
$\R$ and let $\cB:=\{f_i:i\in\N_0\}\cup\{g_j:j\in\N\}$ be a good
basis. Let $\gamma$ be the genus of $H(\sigma)$ (or
equivalently the genus of $H(\rho)$) and 
$\Lambda_{\sigma}$
its largest gap. For a pair of non-negative integers $\ell,m\in\N_0$,
set $a$ to be the (only) integer such that
   $$
m_a\leq m< m_{a+1}
   $$
and let us consider the set
  $$
\R_\ell^m=\{ h\in\R:\text{$\rho(h)\leq \ell$ and $\sigma(h)\leq
m$}\}\, .
  $$
   \begin{proposition}\label{prop6.1}
$\R_\ell^m$ is a vector subspace of $\R.$ Furthermore$,$ if
$m\ge\Lambda_{\sigma}$ then
  $$
\R_\ell^m=\langle f_0,\ldots,f_\ell,g_0,g_1,\ldots,g_a\rangle\, ,
  $$
where $g_0=f_0=1.$ In this case $\dim(\R_\ell^m)=\ell+m+1-\gamma.$
   \end{proposition}
Let $*$ denote the product in $\F^n$ defined by the coordinatewise
multiplication, and let $\varphi:\R\rightarrow \F^n$ be a morphism of
$\F$-algebras. Let $m$ be an integer such that $m\ge\Lambda_{\sigma}$
and $\varphi(\cup_\ell\R_\ell^m)=\F^n$. We define the codes
   \begin{equation}\label{eq6.1}
E_\ell^m:=\varphi(\R_\ell^m)\, \quad\text{and}
\quad C_\ell^m:=(E_\ell^m)^{\perp}\, .
   \end{equation}
Note that, since $\varphi(\cup_\ell\R_\ell^m)=\F^n$,
there exists $L$ such that $E_0^m\subseteq
E_1^m\subseteq\ldots\subseteq E_L^m=\F^n$ and hence
$C_0^m\supseteq C_1^m\supseteq\ldots\supseteq C_L^m=(0)$.
   \begin{example}\label{example6.1} (Continuation of Example
\ref{example5.1}) Let $\cX$ be a curve of genus $\gamma$ over $\F$ and
let $Q_1,Q_2$ be two rational points. Let $\R=\R(Q_1,Q_2)$ and
$\rho,\sigma$ be the n-weights associated to the points $Q_1,Q_2$
respectively. Since $\R_\ell^m=\cL (\ell Q_1+m Q_2)$, if we take a
divisor $D=P_1+\dots+P_n$, sum of $n$ distinct rational points on
$\cX$ and $\varphi=ev$, the evaluation at these points, we obtain the
codes $E_\ell^m=C_{\cL}(\cX,D,\ell Q_1+m Q_2)$ and
$C_\ell^m=C_{\Omega}(\cX,D,\ell Q_1+m Q_2)$.
   \end{example}
        
The dimension of $E_\ell^m$ and $C_\ell^m$ depends on the dimension of
the subspaces $\R_\ell^m$ and the morphism $\varphi$. With regard to their
minimum distances, we shall show a bound on the minimum distance of
$C_\ell^m$, analogous to the order bound in \cite[Section 4]{h-vl-p}.
   
           \subsection{The n-order bound on the minimum 
distance}\label{6.2}

For a vector $\by\in \F^n$ and $i,j=0,\ldots,L$, let us consider the
two-dimensional syndromes
   $$
s_{ij}(\by)=(\bh_i*\bh_j)\cdot\by\, ,
   $$
where $\bh_t=\varphi(f_t)$. The {\em matrix of syndromes} of $\by$ is
$S(\by)=(s_{ij}(\by))_{i,j=0,\ldots,L}$.
    \begin{proposition}\label{prop6.2}\quad $\mbox{\rm
wt}(\by)\geq\mbox{\rm rank} (S(\by)).$
   \end{proposition}
   \begin{proof}
Analogous to \cite[Lemma 4.7]{h-vl-p}.
   \end{proof}
For a nonnegative integer $s$, set
   \begin{align*}
\Sigma(s) & = \max\{\sigma(f_0),\ldots,\sigma(f_s)\}\, ,\quad\text{and}\\
N_\ell^m  & = \{ (i,j)\in\N_0^2 : \text{$i+j=\ell+1$
and $\sigma(f_i)+\Sigma(j)\leq m$}\}\, .
   \end{align*}
Note that for all $i,j$ it holds that
$\rho(f_if_j)=\rho(f_i)+\rho(f_j)$. Thus, if $(i,j)\in N_\ell^m$ then
$f_if_j\in \R_{\ell+1}^m\setminus\R_{\ell}^m$.
    \begin{proposition}\label{prop6.3}
Write $N_\ell^m=\{ (i_1,j_1),\ldots,(i_t,j_t)\}$ ordered in
increasing lexicographical order$.$ Then
   \begin{enumerate}
\item $i_1<\ldots<i_t$ and $j_1>\ldots>j_t;$
\item If $\by\in C_\ell^m$ and $u<v$, then $s_{i_uj_v}(\by)=0;$
\item If $\by\in C_\ell^m\setminus C_{\ell+1}^m,$ then
$s_{i_uj_u}(\by)\neq 0.$
    \end{enumerate}
    \end{proposition}
    \begin{proof} (1) Note that $i_u+j_u=\ell+1$.

(2) Since $j_v<j_u$ then
$\rho(f_{i_u})+\rho(f_{j_v})<\rho(f_{i_u})+\rho(f_{j_u})=\ell+1$ and
$\sigma(f_{i_u})+\sigma(f_{j_v})\leq \sigma(f_{i_u})+c(j_u)\leq m$. Thus
$f_{i_u}f_{j_v}\in R_\ell^m$, hence $\bh_{i_u}*\bh_{j_v}\in E_\ell^m$
and $(\bh_{i_u}*\bh_{j_v})\cdot\by=0$.

(3) Since $f_{i_u}f_{j_u}\in R_{\ell+1}^m\setminus R_\ell^m$, then
$f_{i_u}f_{j_u}=\lambda f_{\ell +1}+f$ with $\lambda\neq 0$ and
$\rho(f)\leq\ell$. Furthermore, since $m\ge\Lambda_{\sigma}$ it holds
that $\sigma(f)\leq m$, hence $f\in\R_{\ell}^m$. Then
$\bh_{i_u}*\bh_{j_u}=\lambda\bh_{\ell+1}+\bh$, with $\bh\in E_\ell^m$,
so $(\bh_{i_u}*\bh_{j_u})\cdot\by= \lambda\bh_{\ell+1}\cdot\by\neq 0$.
    \end{proof}
   \begin{corollary}\label{cor6.1} If $\by\in C_\ell^m\setminus
C_{\ell+1}^m,$ then $\mbox{\rm rank} (S(\by))\geq \# N_\ell^m.$
    \end{corollary}
   \begin{proof} The minor obtained from $S(\by)$ by taking the rows
$i_1<\ldots<i_t$ and the columns $j_1>\ldots>j_t$ is nonsingular.
       \end{proof}
   \begin{definition}\label{definition6.1}
The {\em n-order bound} on the minimum distance of $C_\ell^m$ is
defined as
  $$
d_{NORD}(\ell,m):=\min\{\# N_r^m : r\geq \ell \}\, .
  $$
    \end{definition}
As a direct consequence of the above results we have the following.
   \begin{theorem}\label{thm6.1} The minimum distance of the code
$C_\ell^m$ is lower bounded by $d_{NORD}(\ell,m),$ that is
   $$
d(C_\ell^m)\geq d_{NORD}(\ell,m)\, .
   $$
    \end{theorem}
Next we shall give a bound on the cardinality $\# N_r^m$.
   \begin{proposition}
We have $\# N_r^m\ge \# (H(\rho)\cap [1,r+1]).$ In
particular$,$ if $r\geq \gamma$ then $\# N_r^m\ge r-\gamma+1.$
   \end{proposition}
   \begin{proof}
If $i\in H(\rho)\cap [1,r+1]$ then, according to
\ref{propgood} (3), it holds that $\sigma(f_i)=0$, hence $(i,
r+1-i)\in N_r^m$ (because $\Sigma(r+1-i)\leq \Lambda_{\sigma}\leq m$).
Since $H(\rho)$ is a semigroup of genus $\gamma$ then $\#
(H(\rho)\cap [1,r+1])\ge r-\gamma+1$ for $r\ge \gamma$.
  \end{proof}
  \begin{corollary}\label{cor6.2} If $\ell\ge \gamma,$ then
$d(C_\ell^m)\geq d_{NORD}(\ell,m)\ge \ell-\gamma+1.$
  \end{corollary}
As another consequence of the proposition, the computation of
$d_{NORD}(\ell,m)$ only requires the knowledge of a finite number of
terms $\# N_r^m$.
  \begin{corollary}\label{cor6.3}\quad
$d_{NORD}(\ell,m)=\min\{ \# N_\ell^m,\dots,\# N_{\ell+\gamma}^m\}.$
  \end{corollary}
  \begin{proof} Note that $\# N_r^m\leq r+2$ by definition. Thus,
according to the
above proposition, if $r>\ell+\gamma$ then $\# N_r^m\ge\ell+2\ge
\# N_\ell^m$ and hence $\min\{\# N_r^m : r\geq \ell \}$ must be
attained in the set $\{\# N_\ell^m,\ldots,\# N_{\ell+\gamma}^m\}$.
   \end{proof}
  \begin{remark}\label{rem6.1} Note that the n-order bound does not
depend on the good basis chosen. In fact, since (\ref{eq5.1}) is
equivalent to
   $$
\sigma(f_i)=\min\{t\in\N_0: (i,t)\in H(\rho,\sigma)\}\, ,
   $$
each $\# N_r^m$ (and hence $d_{NORD}(\ell,m)$) can be computed, in
finite time, from only the information given by the semigroup
$H(\rho,\sigma)$.
   \end{remark}
   
   \subsection{Performance of the n-order bound}\label{s6.3}
   
Next we study the performance of the obtained bound. To that end we shall 
compare it to the Goppa bound, $d_G(\ell,m):=\ell+m-2\gamma+2$ by means 
of the number
   $$
\Delta(\ell,m):=d_{NORD}(\ell,m)-d_G(\ell,m)\, .
   $$
Remark that when the code $C_\ell^m$ is obtained from two points on an 
algebraic curve, $C_\ell^m=C_{\Omega}(\cX,D,\ell Q_1+m Q_2)$, then its 
minimum distance verifies $d(C_\ell^m)\geq d_{G}(\ell,m)$.

Let $\Lambda_\rho$ and $\Lambda_\sigma$ be the largest gaps of $H(\rho)$ 
and $H(\sigma)$ respectively. Furthermore, let $s$ be the integer defined by
  $$
\sigma(f_{s})=\max\{\sigma(f_i): i\in \N\}=\Lambda_\sigma=c_\sigma-1\ .
 $$
For large values of $m$ the n-order bound is easy to compute. 
  \begin{lemma}\label{lemma6.1} If $m\geq 2\Lambda_\sigma,$ then 
$d_{NORD}(\ell,m)=\ell+2.$ In particular$,$ $\Delta(\ell,m)=2\gamma-m$, 
hence $\Delta(\ell,m)<0$ 
for $m>2\gamma$ and $d_{NORD}(\ell,m)=d_G$ if and only if $m=2\gamma$ (and thus 
$\Lambda_\sigma=\gamma).$
  \end{lemma}
  \begin{proof}
Since $\sigma(f_i)\leq\Lambda_{\sigma}$ and $\Sigma(i)\leq\Lambda_\sigma$ 
for all $i$, $\#N_r^m=r+2$ by hypothesis; thus $d_{NORD}(\ell,m)=\ell+2$ 
and the result follows.
  \end{proof}
Thus, the remaining case to study is 
$\Lambda_\sigma\leq m< 2\Lambda_\sigma$. Write
  \begin{align*}
N_r^m= \{ (0,r+1),(r+1,0)\} \cup \{(i,j)\in\N^2 : i\in A^m_r\cup 
B^m_r\cup C^m_r\}\, ,
   \end{align*}
where
    \begin{align*}
A_r^m &=H(\rho)\cap [1,r], \\
B_r^m &=\{i\in \Gaps(H(\rho))\cap[1,r+1-s]: 
\sigma(f_i)+\Lambda_\sigma\leq m\}\quad\text{and} \\
C_r^m&=\{i\in \Gaps(H(\rho))\cap [r+2-s,r] :  
\sigma(f_i)+\Sigma(r+1-i)\leq m\}.  
    \end{align*}
The following lemma holds true.
   \begin{lemma}\label{lemma6.2} Assume $\Lambda_\sigma\leq 
m<2\Lambda_\sigma$ and let $\ell\geq \Lambda_\rho+s-1$. Then
    \begin{enumerate}
\item $d_{NORD}(\ell,m)=\ell+2-\gamma+\# A_\ell^m;$
\item If $\Lambda_\sigma\geq \gamma +1,$ then $d_{NORD}(\ell,m)<d_G(\ell,m);$
\item $d_{NORD}(\ell,m)=d_G(\ell,m)$ if and only if $\Lambda_\sigma=\gamma.$
   \end{enumerate}
   \end{lemma}
   \begin{proof} (1) We have $r\geq \Lambda_\rho+1$ for $r\geq \ell$; on 
the other hand, $\ell+2-s\geq \Lambda_\rho+1$ and the proof follows 
from the fact that $\# A_r^m$ increases with $r$.

(2) By (1), and since $\# A_r^m\leq m-\Lambda_\sigma$, we have
   $
\Delta(\ell,m)=\gamma+\# A_\ell^m-m\leq \gamma-\Lambda_\sigma .
   $

(3) If $\Delta(\ell,m)=0$, then clearly $\Lambda_\sigma= \gamma$. 
Conversely, if $\Lambda_\sigma=\gamma$, then $\# A_\ell^m=m-\gamma$. 
    \end{proof}
After this lemma, one may expect to obtain $\Delta(\ell,m)>0$ only in the 
case 
    $$
  \Lambda_\sigma\leq m<2\Lambda_\sigma\quad\text{and}\quad \ell\leq 
\Lambda_\rho+s-2\, .
   $$
In fact, this can occur as the next example shows. 
   \begin{example}\label{ex6.1} Suppose $\sigma(f_i)=i$ for 
$i=1,\ldots,\gamma$ (this case can occur on points of the Hyperelliptic 
curve, see \cite{homma}). Then $H(\rho)=\{\gamma+1,\gamma+2,\ldots\}$ and 
$s=\gamma$. Take $\gamma\leq m<2\gamma$ and $\ell\geq \gamma+1$; thus for 
$r\geq \gamma$
   \begin{align*}
\# N^m_r  & = r+2-\gamma +  \\
{} & \#\{i\in [1,r+1-\gamma]\cap\Gaps(H_\rho): i+\gamma\leq m\}+\\
 & {} \# \{i\in [r+2-\gamma,r+1]\cap\Gaps(H_\rho): i+\Sigma(r+1-i)\}\, .
\end{align*}
Since $m<\gamma$,
    $$
\# N^m_r=r+2-\gamma+\min\{r+1-\gamma,m-\gamma\}+c(m,r)\, ,
    $$
where $c(m,r)=0$ if $r+1>m$  and  
$c(m,r)=2\gamma-r-1$ if  $r+1\leq m+1$. 
Thus
  $$
\# N^m_r=\begin{cases} r+m-2\gamma+2 & \text{if $r+1>m$,}\\
  r+2 & \text{if $r+1\leq m$.}
    \end{cases}
   $$
Observe that $\# N^m_\ell=d_G(\ell,m)$ if $\ell+1>m$ and $\# 
N^m_\ell>d_G(\ell,m)$ otherwise. Thus $d_{ORD}(\ell,m)$ is greater than 
$d_G(\ell,m)$ whenever $\gamma\leq \ell< m<2\gamma$.

Finally, for the case $\gamma<m<2\gamma$ and $\ell<\gamma$, a direct 
computation shows that $\# N_r^m=r+2$ and hence the n-order bound on the 
minimum distance improves also on the Goppa bound.
   \end{example}


\begin{thebibliography}{99}

\bibitem{ct} Carvalho C. and Torres F., {\em On Goppa codes and
Weierstrass gaps at several points}, Des. Codes Cryptogr. {\bf 35}(2)
(2005), 211--225.

\bibitem{feng} Feng G.L. and Rao T.R.N., {\em Improved geometric Goppa
codes part I, basic theory}, IEEE Trans. Inf. Theory {\bf 41}(6)
(1995), 1678--1693.

\bibitem{garcia-lax1} Garcia A. and Lax R., {\em Goppa codes and
Weierstrass gaps}, Lecture Note in Math., Springer-Verlag,
Berlin-Heildelberg, {\bf 1518}, 33--42, 1992.

\bibitem{garcia-lax2} Garcia A., Kim S.J. and Lax R., {\em Consecutive
Weierstrass gaps and minimum distance of Goppa codes}, J. Pure Appl.
Algebra {\bf 84} (1993), 199--207.

\bibitem{ord1} Geil, O. and Pellikaan R., {\em On the structure of order 
domains}, Finite Fields and their Applications {\bf 8} (2002), 369--396.

\bibitem{goppa1} Goppa, V.D., {\em Codes associated with divisors},
Problems Inform. Transmission {\bf 13} (1977), 22--26.

\bibitem{goppa2} Goppa, V.D. ``Geometry and Codes", Mathematics and
its Applications, vol 24, Kluwer, Dordrecht (1991).

\bibitem{h-vl-p} H\o holdt, T., van Lint J.V. and Pellikaan R., {\em
Algebraic Geometry Codes}, Handbook of Coding Theory, eds. V. Pless
and W.C. Huffman, 871--961, Elsevier, 1998.

\bibitem{homma} Homma, M., {\em The Weierstrass semigroup of a pair of
points on a curve}, Arch. Math. {\bf 67} (1996), 337--348.

\bibitem{homma-kim1} Homma, M. and Kim, S.J., {\em Goppa codes with
Weierstrass pairs}, J. Pure Appl. Algebra {\bf 162} (2001), 273--290.

\bibitem{homma-kim2} Homma, M. and Kim, S.J., {\em Toward the
determination of the minimum distance of two-point codes on a
Hermitian curve}, Des. Codes Cryptogr., to appear.

\bibitem{homma-kim3} Homma, M. and Kim, S.J., {\em The two-point codes
on a Hermitian curve with the designed minimum distance}, Des. Codes
Cryptogr., to appear.

\bibitem{homma-kim4} Homma, M. and Kim, S.J., {\em The two-point codes
with the designed distance on a Hermitian curve in even
characteristic}, preprint.

\bibitem{kim} Kim, S.J., {\em On the index of the Weiertrass semigroup
of a pair of points on a curve}, Arch. Math. {\bf 62} (1994), 73--82.

\bibitem{matsumoto} Matsumoto, R., {\em Miura's generalization of
one-point AG codes is equivalent to H\o holdt, van Lint and
Pellikaan's}, IEICE TRANS. FUNDAMENTALS {\bf E82}-A(10) (1999),
2007--2010.

\bibitem{gretchen3} Matthews, G., {\em Weierstrass pairs and minimum
distance of Goppa codes}, Designs Codes Cryptogr. {\bf 22} (2001),
107--221.

\bibitem{gretchen1} Matthews, G., {\em The Weierstrass semigroup of an
$m$-tuple of collinear points on a Hermitian curve} (A. Poli, H.
Stichtenoth Eds.) Fq7 2003, LNCS {\bf 2948}, 12--24, 2004.

\bibitem{gretchen2} Matthews, G., {\em Weierstrass semigroups and
codes from a quotient of the Hermitian curve}, preprint.

\bibitem{gretchen-survey} Matthews, G., {\em Some computational tools
for estimating the parameters of algebraic geometry codes},
Contemporary Mathematics {\bf 381} (2005), 19--26.

\bibitem{ord2} Pellikaan R., {\em On the existence of order functions}, 
Journal of Statistical Planning and Inference {\bf 94} (2001), 287--301. 

\bibitem{ercilio} Silva, E., ``Func\~oes Ordens Fracas e a Dist\^ancia
M\'{\i }nima dos C\'odigos de Goppa Geom\'etricos", Tese (Doutorado),
http:\slash\slash libdigi.unicamp.br/document/?code=vtls000333125,
IMECC-UNICAMP, Cx. P. 6065, 13083-970, Campinas SP-Brazil.

\bibitem{tsf-vla-z} Tsfasman M.A., Vl\u{a}du\b{t} S.G. and Zink T.
{\em Modular curves, Shimura curves and Goppa codes, better than
Varshamov-Gilbert bound}, Math. Nachr. {\bf 109} (1982), 21--28.

   \end{thebibliography}
   \end{document}